\documentstyle[prl,aps,preprint,epsfig,tighten]{revtex}

\begin{document}
\draft

\title{Resonance-induced effects in photonic crystals}

\author{Alexander 
Moroz\thanks{http://www.amolf.nl/external/wwwlab/atoms/theory/} and  
Adriaan Tip${}^*$}
\address{FOM-Instituut voor  Atoom- en Molecuulfysica, Kruislaan 407, 
1098 SJ Amsterdam,\\ The Netherlands}
\address{
\begin{abstract}
For the case of a simple face-centered-cubic photonic crystal of 
homogeneous dielectric spheres, we examine to what extent
single-sphere Mie resonance frequencies are related to band gaps
and  whether the width of a gap can be enlarged due to nearby 
resonances. Contrary to some suggestions, no spectacular effects
may be expected. When the dielectric constant of the spheres $\varepsilon_s$ 
is greater than the dielectric constant $\varepsilon_b$ of the 
background medium,  then for any filling fraction $f$ there exists a 
critical $\varepsilon_c$ above which the lowest lying Mie resonance 
frequency falls inside the lowest stop gap 
in the (111) crystal direction, close to its midgap frequency.
If $\varepsilon_s <\varepsilon_b$, the correspondence
between Mie resonances and both the (111) stop gap and a full gap 
does not follow such a regular pattern. If the Mie resonance frequency 
is close to a gap edge, one can observe a resonance-induced widening of 
a relative gap width by $\approx 5\%$. \vspace*{1.4cm} \\
\end{abstract}
}

\address{
({\bf J. Phys.: Condens. Matter 11, 2503-2512  (1999)})
\vspace*{1.4cm}}
\maketitle


\newpage
\narrowtext

\section{Introduction}
Photonic crystals, i.e., dielectrics with spatial periodicity,
have triggered much interest recently \cite{Y,Jo,HCS,YGL}.
One can picture a photonic crystal as a periodic arrangement of
dielectric scatterers. For example, its dielectric constant 
$\varepsilon({\bf r})$ equals  $\varepsilon_s$ for ${\bf r}$  
inside the scatterer and $\varepsilon_b$ otherwise. 
Similarly to the case of an electron moving in a periodic potential,
a photon traveling in a photonic crystal encounters
a periodicaly changing dielectric constant. As a result,
a gap can open in the electromagnetic wave spectrum, independent of 
its polarization and direction of propagation \cite{HCS,YGL}. 
In a given frequency interval the density of states (DOS) can either 
be reduced down to zero (photonic band gap) or enhaced with respect to 
its vacuum value. Such a change in the DOS affects many physical 
quantities. The most transparent is the change in the spontaneous 
emission (SE) rate of embedded atoms and molecules. This can be demonstrated
already at a relatively low refractive index contrast 
($\approx 1.15$) \cite{MWL}.
Suppression of the SE may have applications 
for semiconductor lasers, solar cells, heterojunction bipolar 
transistors, and thresholdless lasers \cite{Y}. On the other hand, 
the enhancement of the SE is a way to create  new sources of light 
for ultra-fast optical communication systems \cite{RK}. 
Unlike conventional (electronic) crystals, photonic crystals are 
essentially man-made structures and their parameters can be changed at will. 
There is a common belief that in the near future such systems will allow 
us to perform many functions with light that ordinary crystals do 
with electrons.

Thus far, the main emphasis in the study of 
photonic crystals has  been on calculating the band structure
\cite{HCS,SHI,BSS,MS}. 
Let $f$ be the scatterer filling fraction, i.e., the 
volume of the scatterer(s) in the unit cell per unit cell volume. 
Once $f$ is fixed, the spectrum is only a function of the dielectric 
contrast 
$\delta=\max (\varepsilon_s/\varepsilon_b,\varepsilon_b/\varepsilon_s)$.
In our paper we pose the question of to what extent the single 
sphere resonance frequencies are related to band gaps, whether a gap
width  can be enlarged due to nearby resonances, and 
what other effects, if any, single-scatterer resonances may have on the band 
structure and properties of a  photonic crystal. 

In the following,  we focus on the case of a simple 
face-centered-cubic (fcc) photonic crystal of homogeneous spheres.
There are at least two reasons to consider this case.
Firstly, single-sphere resonances, known also as Mie resonances 
\cite{Mie}, are well understood and an analytic solution exists for them. 
In each angular-momentum channel characterized
by the angular-momentum number $l$ a single sphere has 
an infinite number of Mie's resonances. The properties of 
Mie's resonances are discussed  in many monographs 
(see, for example, \cite{BH}) and we only emphasize the following
ones. The sharpness of a Mie resonance decreases with increasing
$\sigma r_s$, where $r_s$ is the sphere radius,
$\sigma=\omega\sqrt{\varepsilon_b}/c$, $\omega$ is the frequency,
and $c$ is the speed of light in vacuum. On the other hand, 
reflecting the centrifugal barrier 
increasing with $l$ as $l(l+1)$, the sharpness increases with $l$.
The spacing between resonances in the higher frequency range is 
accounted for by the resonance condition $\sigma r_s= m \pi /2$, where 
$m$ is an integer (if $\varepsilon_b=1$).

Secondly, fcc structures of homogeneous spheres are among 
the most promising candidates to achieve a full photonic gap at optical 
and near-infrared frequencies.
Indeed, in this frequency range one often uses collodial systems of 
microspheres  which can self-assemble into 
three-dimensional fcc crystals with excellent long-range periodicity 
\cite{Pier,WV}. This long-range periodicity gives rise to the strong 
optical Bragg scattering, clearly visible by the naked eye, and 
already described in 1963 \cite{LKW}. 
Both the case of ``dense'' spheres 
($\varepsilon _{s}>\varepsilon _{b}$) \cite{WV} and 
``air'' spheres ($\varepsilon _{s}<\varepsilon _{b}$) \cite{IP}  
can be realized experimentally.

The outline of our paper is as follows. In the next section,
main features of the method  that is used to calculate 
the spectrum of electromagnetic waves in periodic structures are discussed.
We also discuss the differences between and similarities
of the electronic and electromagnetic bands and give their
rough classification. In section \ref{sec:result}
we summarize  our numerical results on the effects induced by
resonant scattering in a simple fcc crystal of dielectric spheres.
Contrary to the suggestions made previously in the literature
\cite{Jo}, no spectacular effects may be expected.
Finally, in section \ref{sec:concl} our conclusions are presented.

\section{Resonance scattering and band structure of photonic crystals}
\label{sec:ressc}
Multiple scattering of classical waves in the presence of resonant 
scatterers was first studied in a disordered medium. There 
resonances cause a scattering delay due to 
the storage of wave energy inside a single scatterer, resulting in a 
sharp decrease (depending on the filling fraction)
of the transport velocity $v_E$ of light \cite{vATL}. 
The discussion of resonance-induced effects in an ordered medium 
has been initiated by John \cite{Jo}.
He suggested that the coinciding resonance and 
Bragg scattering is the most favourable condition for opening a gap in the 
spectrum. In the case of spheres this
leads to $f=1/(2\sqrt{\delta })$ \cite{Jo}. Later on,
Zhang and  Satpathy \cite{ZS} noticed that a pseudogap in the
band structure of an fcc lattice of dense spheres
corresponds to a Mie resonance. Recently,
Ohtaka and Tanabe \cite{OT1}, using the KKR method described earlier
in \cite{Mo}, made an attempt to relate the 
Mie resonances to the photonic bands. They paid attention to 
flat bands which correspond to ``heavy photons" in analogy to their 
electronic analogue, heavy fermions. For heavy photons, the group velocity
can be as low as $\sim c/100$ \cite{OT1}. 
The role of Mie resonances 
in bonding of spheres in photonic crystals was investigated 
by Antonoyiannakis and  Pendry \cite{AP}.

In our paper, we shall attempt to investigate the 
complementary relation to that discussed by Ohtaka and Tanabe \cite{OT1},
namely, to what extent the single 
sphere resonance frequencies are related to band {\em gaps}, a question
asked first by John \cite{Jo}.
In obtaining his result, John \cite{Jo} used a simplified picture
in which, regardless of the propagation and polarization, the photon
always encounters precisely the same periodic $\varepsilon({\bf r})$
variation, resulting in a one-dimensional Kronig-Penney's (KP) model 
\cite{KP}. However, this approximation is only partially
justified in the short-wavelength limit when light in a sphere 
behaves like a field in a slab with thickness $2r_s$.
For real crystals the situation is different
and the above coincidence condition can only be met
in restricted  regions  on the surface of the Brillouin zone.
Also, with increasing dielectric contrast $\delta$, dispersion
curves $\omega=\omega({\bf k})$ become nonlinear and
the diffraction condition is modified as compared to 
the Bragg case and fulfilled at lower frequency \cite{WV}.
Moreover, the model neglects polarization dependent effects.

However, it is quite difficult to relax the simplifying assumptions 
of the one-dimensional KP model \cite{KP}
made by John \cite{Jo}. 
In our discussion of resonance-induced effects in
photonic crystals we shall employ the machinery of the on-shell 
multiple-scattering theory 
(MST) (see \cite{Bb} for electrons and \cite{Mo} for photons).
Note that the on-shell MST is already required in the one-dimensional
KP model if one wants to go beyond the dispersion relation
and obtain Green's function or the local density of states \cite{AMp}. 
The unique feature of the on-shell MST is that, for nonoverlapping 
(muffin-tin) scatterers \cite{Bb,LS}  (the present situation),
it disentangles single-scattering and 
multiple-scattering effects  (see \cite{MT} for a recent discussion).
For example, the total T-matrix per unit cell can be written in the  
form \cite{Bb,Mo,LS,Zi}
\begin{equation} 
T = 1/(R^{-1}-B) =R /(1-BR).
\label{ttot}
\end{equation} 
Here all quantities are ordinary matrices. $R$ stands for the 
on-shell reaction matrix (also known as the k-matrix \cite{LS}) of 
the scattering sphere (of all scatterers 
inside the lattice unit cell in the case of a complex lattice).
It is diagonal in the angular-momentum basis
and can be written as $R_{LL'}=-\delta_{LL'}\tan \eta_L/\sigma$,
where $\eta_L$ is a  phase shift. Here
$L$ is  a composite index which labels all the spherical
harmonics in the irreducible representation of the rotation group 
characterized by the principal angular-momentum number $l$ and,
in the case of electromagnetic waves,  it carries an additional
polarization dependent index \cite{Mo}.
$B=B(\sigma,{\bf k})$ in Eq. (\ref{ttot}) is a matrix of so-called 
structure constants which accounts for the periodicity of the lattice.
It depends on $\sigma$ and the Bloch momentum ${\bf k}$.
The R-matrix is singular at Mie resonance frequencies
and $B(\sigma,{\bf k})$ is singular whenever 
$\sigma^2=({\bf k}+{\bf K}_n)^2$,
where ${\bf K}_n$ is a vector of the dual lattice.

The exact eigenmodes of a crystal are determined by  poles of the 
total T-matrix, the latter being the zeros of the determinant of a hermitian
matrix $R^{-1}-B$,
\begin{equation}
\det (R^{-1}-B)=0.
\label{kkr}
\end{equation}
Equation (\ref{kkr}) is the familiar Korringa-Kohn-Rostocker (KKR)
equation \cite{KKR} in band structure theory.
Recently, we have succesfully used this approach 
to calculate the band structure of electromagnetic waves
in a simple fcc lattice 
of dielectric spheres \cite{MS}, to establish a simple
analytic formula describing the width of the lowest lying
stop gap in the $(111)$ crystal direction 
(the L point of the Brillouin zone \cite{Kos})
\cite{Mo1}, and to calculate the properties of the
local DOS in one dimension \cite{AMp}.
In general, the higher frequency, the higher 
the value of  $l_{max}$ is needed to ensure convergence. 
In order to reproduce the first
band and the linear part of the spectrum, $l_{max}=1$  is enough. 
In general, the size of a secular equation is reduced by almost a factor $10$
compared with that in the plane-wave method \cite{HCS,SHI,BSS,ZS} which 
customarily requires well above a thousand plane waves. 
The precision of the elements
of the secular equation is determined by the standard Ewald 
summation \cite{KKR} which yields structure constants up to six digits.

At first sight, there seem to be large differences
in band formation between the electrons and electromagnetic waves.
In the case of electrons, in the medium between two atoms, waves
are evanescent in nature, while electromagnetic wave propagates
unattenuated between two scatteres.
However, electrons are strongly interacting with each other.
If the interactions are taken into account, an effective-Hamiltonian
single-electron picture emerges of a near-free electron with a positive 
energy which moves unattenuated between two scatteres
like the electromagnetic wave does \cite{Zi}. Therefore, the same 
principles apply to the classification of 
the electronic and the photonic bands.
For electrons, a rough classification of the bands can be obtained 
if the singularities of $R$ and $B$ are well separated \cite{Zi}.
Thus if $R(\sigma)$ is sufficiently small near a singularity of 
$B(\sigma,{\bf k})$, where $\sigma=\sqrt{E}$ and $E$ is the
electron energy, a band is formed with the dispersion relation 
\begin{equation} 
\omega ({\bf k})\approx |{\bf k}+{\bf K}_n| + R(\sigma).
\end{equation}
In more intuitive terms, the formation of such a band 
results from the formation of standing waves in a crystal
and it is appropriate to call such a band the Bragg band.
The second type of band can form near a singularity of 
$R$ at $\omega_0$, $R=\Gamma/(\omega^2-\omega^2_0)$. 
This is, for example, the case in the transition and noble metals.
If $\Gamma$ and  $B(\sigma,{\bf k})$ in the vicinity 
of $\omega_0$ are {\em sufficiently} small,
a (usually very narrow) resonance band is formed with the 
dispersion  relation \cite{Zi}
\begin{equation} 
\omega ({\bf k}) \approx [\omega_0+\Gamma B(\sigma_0,{\bf k})]^{1/2}.
\label{rbw}
\end{equation}
Eq. (\ref{rbw}) is in agreement with the observation made
in \cite{OT1} that the resonance-band width is {\em comparable} with 
the lifetime broadening of the DOS profile for a 
single sphere. In more intuitive terms,  formation of the resonance band
can be understood as resulting from the broadening of 
individual resonances when they start to feel the presence of 
each other, similar to the formation of the electronic bands 
from individual atomic levels in the tight-binding
limit \cite{Zi}. Such usually very narrow resonance bands describe
``heavy photons". 

This shows that one can associate with a resonance a (resonance) band.
Is it possible to associate with a resonance a gap?
The answer is yes. However,  a hybridization of bands must 
take place. We speak of hybridization if the singularities of $R$ and 
$B$ cannot be well separated and  neither a pure Bragg nor a pure resonance 
band is  formed. Under certain conditions the two bands can 
hybridize in such a way as to create a gap over the approximate
energy range of the original unhybridized resonance band.
An example of such a hybridization is provided by transition metals 
with characteristic $d$ (or $f$) resonance which 
couples with extended band states by tunneling \cite{Har}.
In the latter case a broad $s-p$ band hybridizes with a narrow $d$
band in such a way as to create a gap over the approximate
energy range of the original unhybridized $d$ resonance band \cite{Har}.

In the following section, we investigate the hybridization of
photonic bands in an fcc lattice of dielectric spheres.
Although the same principles apply to the classification
of the photonic bands as to those for electrons,
is does not mean that the respective band structures
are qualitatively similar. There is significantly difficult to 
open a gap in the spectrum of electromagnetic waves than in the case 
of electrons. Moreover, a gap often does not 
open between the lowest lying bands, as in the case of electrons,
but in an intermediate region. The origin of this difference
lies in a different behaviour of $R$ and $B$ and 
can be rather easily understood. Indeed, 
in the tight-binding picture of band formation \cite{Zi},
individual bands results from the broadening 
of corresponding atomic levels when the atoms start to feel the presence 
of each other.  The largest gap between atomic levels is
between the lowest-lying energy levels. Therefore, for a lattice
of atoms, one expects to find a gap essentially between the first
and the second energy band, with the gap  between
higher bands scaling down to zero. On the other hand, for a dielectric 
scatterer and Maxwell's equations, bound states are absent. 
They are replaced by resonances. Moreover, if the wavelength is small 
compared to the size of the spheres, one can use geometric optics, while in the
opposite limit of long wavelengths, the Rayleigh approximation applies.
In neither case does a gap open in the spectrum.
Therefore, if a gap is present in the spectrum, it should be
in the intermediate region between the two limiting cases
(see, however, the case of a diamond lattice (\cite{HCS}, figure 2), 
which is a complex lattice).
The same applies to the localization
of light \cite{Jo} which is also expected at some intermediate 
frequencies.

Since opening of a gap in the spectrum of electromagnetic waves
in much difficult than in the electronic case, one expects also
that hybridization will be weaker in the former case.
These expectations are confirmed in the  following section.

\section{Results for an fcc lattice of dielectric spheres}
\label{sec:result}
In our case of an fcc crystal of homogeneous spheres, 
we looked for a correspondence between Mie resonance frequencies
and (i) the lowest lying stop gap in the (111) crystal direction
(the L point of the Brillouin zone \cite{Kos})
in the case of both  ``dense'' spheres ($\varepsilon_s>\varepsilon_b$) 
and  ``air'' spheres ($\varepsilon_s <\varepsilon_b$),
(ii) the full gap in the case of ``air'' spheres.
(Note that there is no full gap in the case of ``dense'' spheres
\cite{SHI,MS}.)
Apart from numerous experimental 
data now available \cite{WV,YG}, there are at least two other
reasons to chose the L-gap. First, the width of the first stop gap 
often takes on its  maximum at the L point and, second, experimental techniques 
make it possible to grow collodial crystals such that the L direction 
corresponds to normal incidence on the crystal surface.

\begin{figure}[tbp]
\begin{center}
\epsfig{file=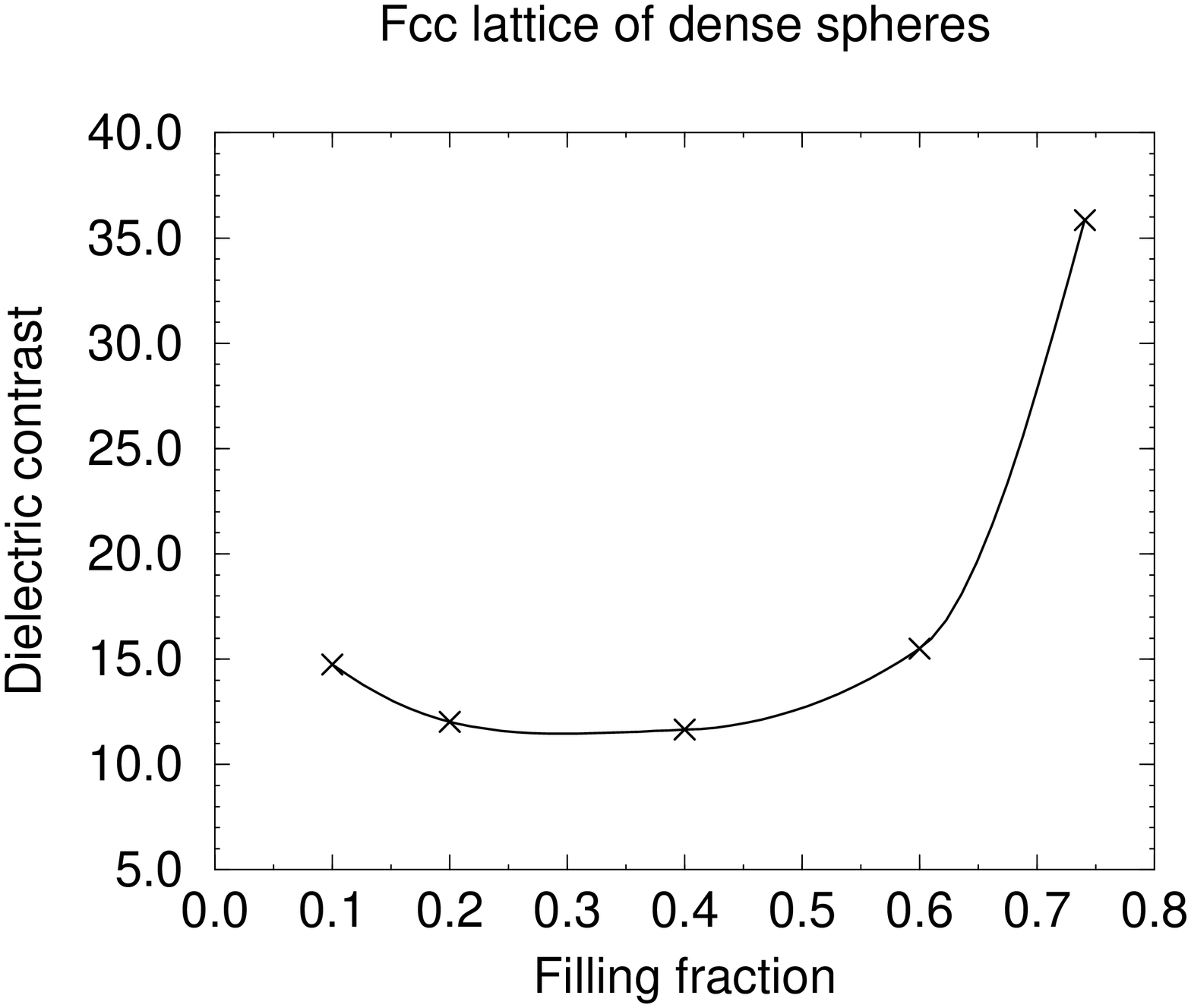,width=10cm,clip=0,angle=-90}
\end{center}
\caption{The value of the dielectric contrast 
at which the lowest Mie resonance enters the L gap versus filling
fraction in the case of an fcc lattice of homogeneous dense
spheres.  
}
\label{rvff}
\end{figure}

In the case of ``dense'' spheres we found that for sufficiently 
high dielectric contrast $\delta$ 
the L-gap can be associated with the 1M1 resonance.
(Our  notation $lAn$ for a Mie resonance is such that
$l$ is the angular momentum, $A$ stands for either 
electric (E) or magnetic (M) mode, and $n$ is the order
of the resonance with increasing frequency in a given $lA$ channel.)
For all filling fractions one finds that as $\delta$ 
increases over a critical value (see Figure \ \ref{rvff}),
the 1M1 resonance ``descends'' from above to the L-gap and 
stays inside it close to the midgap frequency. We verified this
behaviour for $\delta$ up to $100$.

In the case of ``air'' spheres the hybridization of bands 
is much less pronounced compared to the case of ``dense'' spheres. 
We observe a correspondence between the L-gap and a Mie resonance
only for particular filling fractions and dielectric contrasts.
Filling fraction $f\approx 0.5$ or higher is required for 
hybridization to occur. However, for a given $f$, the required 
dielectric contrast for the onset of  hybridization  can be as little as
half the value for the dense sphere case.
For $f=0.6$ one finds the 1E1 resonance trapped inside the L-gap 
already for  $\delta \geq 8$. In the close-packed case
the lowest lying 1E1 resonance descends actually below the L-gap
and, if  $\delta \geq 16$, a hybridization occurs at the frequency 
corresponding to the second, 1M1, resonance.

\begin{figure}[tbp]
\begin{center}
\epsfig{file=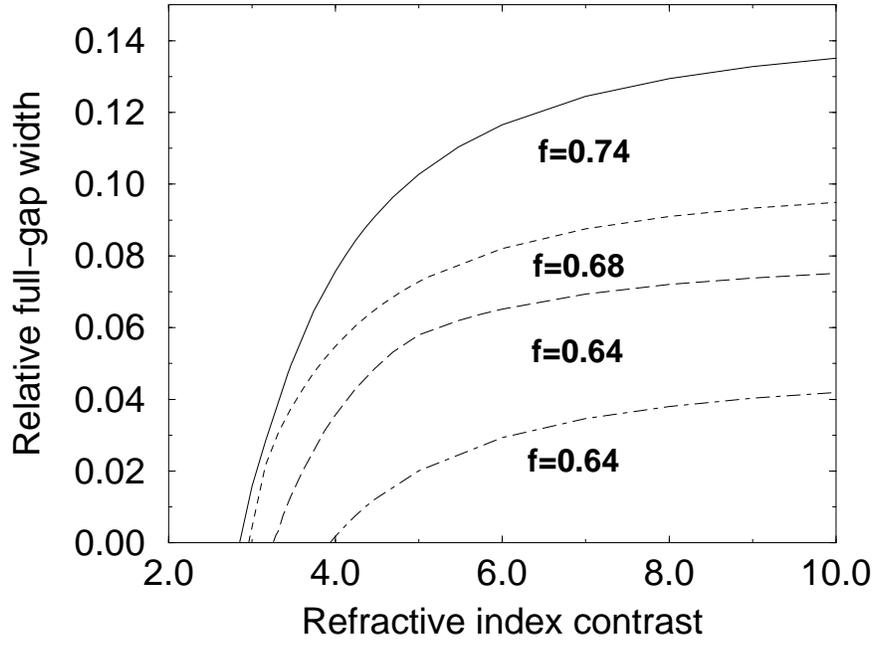,width=10cm,clip=0,angle=-90}
\end{center}
\caption{The relative width $\Delta^r$ of the full gap 
(the full gap width divided by the midgap frequency) 
in the case of an fcc lattice of homogeneous air spheres as a function 
of the refractive index contrast $\sqrt{\delta}$
for different filling fractions $f$.}
\label{fgrww}
\end{figure}

Hybridization in the case of the lowest L-gap
is only  partial since the gap does not extend
over the whole Brillouin zone.
Let us, therefore, look at the full band gap which
can be opened only in  the air sphere
case (one can open just a single full gap here  \cite{SHI,MS}). 
In contrast to that of the L-gap, the opening of the full gap requires
a certain threshold dielectric contrast which rapidly increases
as $f$ decreases from the close-packed case \cite{SHI,MS}
(see figure \ref{fgrww}).
Hybridization follows the irregular pattern seen in the case of the L-gap.
For example, for $f=0.6$ hybridization occurs
if $\delta\geq 36$  where the  2M1 resonance descends from 
above to the
full gap, soon followed by the 3E1 resonance (2M1 $<$ 3E1). 
Both resonances seem to be locked inside the full gap
(at least up to  $\delta=100$).
Interesting behaviour is found for $f=0.64$.
First, the full gap  opens at
$\delta \approx 10.6$ between the closely lying 2M1 and 3E1  
resonances.
As  $\varepsilon_b$ increases, the 3E1 resonance
moves across the gap and stays closely below the lower edge of 
the full gap.
Hybridization around the 3E1 resonance occurs only
for  $\delta=(11.5,13)$. Outside this interval of $\delta$ 
no resonance is inside the full gap.
The case in which $f=0.68$ shows an anomalous behaviour:
no resonance frequency is inside the full gap or close to the the full 
gap edge.
On the other hand, in the  close-packed case ($f\approx 0.74$) 
one can find  three closely lying resonances, namely, 3M1,  4E1, and 1E2 
(3M1 $<$  4E1 $<$  1E2), inside the 
full gap. 3M1 descends to the full gap around $\varepsilon_b=12$,
soon followed by the 4E1 and 1E2 resonances. For 
$\varepsilon_b>16$ all three resonances are already 
inside the full gap and seem to remain trapped there 
(at least up to $\varepsilon_b=100$).

\begin{figure}[tbp]
\begin{center}
\epsfig{file=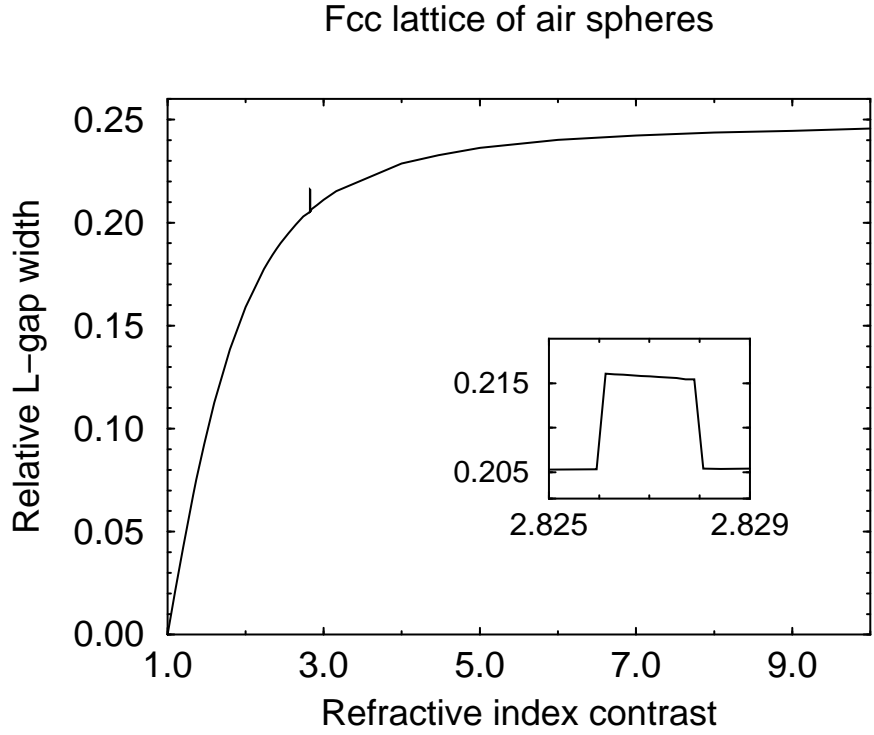,width=10cm,clip=0,angle=-90}
\end{center}
\caption{The relative L-gap width $\Delta_L^r$
(the L-gap width divided by the
midgap frequency) for an fcc lattice of air spheres
with the sphere filling fraction $f=0.6$ plotted against the 
refractive index contrast $\sqrt{\delta}$.
Around $\delta=7.99$, the 1E1 resonance crosses the upper edge of
the L-gap which results in a local enhancement of the   
relative L-gap width, which is shown as a very narrow peak
in the main figure. The inset shows the detailed view of the peak. 
}
\label{rlrgap}
\end{figure}

Apart from the hybridization, another intriguing issue is that of
whether the width of a gap can be enlarged due to the Mie resonances 
as  the one-dimensional KP model suggests \cite{Jo}.
Once a resonance was inside a gap we did not find any 
significant effect on the gap width.
A good illustration of this is the behaviour of the relative
width $\Delta^r$ of the full gap
(the full gap width divided by the midgap frequency)
at $f=0.68$ on the one hand and $f=0.6,\, 0.64$ and $0.74$ 
(close-packed) on the other hand (see figure \ref{fgrww}).
In the  first case, the gap does not correspond to any Mie resonance,
while in the second case, there are up to three Mie resonances
inside the gap. Had there been some effect of a Mie resonance
on the gap width, one would have observed either 
a sudden increase in $\Delta^r$ after hybridization sets in,
or an anomalously small gap width for 
 $f=0.64$ (if $\delta > 13$) and $f=0.68$. 
Instead, one sees a monotonical
increase of $\Delta^r$ as $\delta$ and $f$ increase.
However, something different may happen if a Mie resonance frequency 
is close to a gap edge, before hybridization sets in.
Under certain circumstances  one can observe a resonance-induced
widening of a relative gap width by up to $ 5\%$.
Figure\ \ref{rlrgap} shows the relative L-gap  width $\Delta_L^r$
for air spheres and $f=0.6$. Around $\delta=7.99$ 
the 1E1 resonance crosses the upper edge of
the L-gap, which results in a local enhancement of 
$\Delta_L^r$ by $\approx 5\%$.
As a function of $\delta$, this widening
occurs within the very narrow interval of $\delta\in(7.986,7.998)$
where it shows a flat peak.
In the air-sphere case one can be sure that this widening of a gap
can be entirely attributed to a resonance. 
As shown recently \cite{Mo1}, 
the lowest lying L-gap for an fcc lattice of dielectric spheres 
can be understood in terms of simple quantities, 
namely, the volume averaged dielectric constant, 
$\overline{\varepsilon} = f\varepsilon_s +(1-f)\varepsilon_b,
$
the volume averaged $\varepsilon^2({\bf r})$,
$\overline{\varepsilon^2}=
[f\varepsilon_s^2 +(1-f)\varepsilon_b^2],
$
and the effective dielectric constant $\varepsilon_{e\!f\!f}$.
The effective dielectric constant can be well approximated 
\cite{MS,DCH} by Maxwell-Garnett's formula \cite{MG},
\begin{equation}
\varepsilon_{e\!f\!f} \approx \varepsilon_b \,
(1+ 2\, f\alpha)/(1- f\alpha)
\label{maxgsp}
\end{equation}
where, for a homogeneous sphere, 
the polarizability factor
\begin{equation}
\alpha=(\varepsilon_s-\varepsilon_b)/(\varepsilon_s+2\varepsilon_b).
\end{equation}
Then the absolute L-gap width $\triangle_L$ is approximated 
within 3-6\% (depending on $f$) by the formula \cite{Mo1}
\begin{equation}
\triangle_L =  \left. C(f) \left(\sqrt{
    \overline{\varepsilon^2}} -\varepsilon_{e\!f\!f}\right)^{1/2}
\right/ \bar{\varepsilon}
\label{best}
\end{equation}
where
\begin{equation}
C(f)=C_{0} + 0.14\, f\,(2 f_{m} -f)/f_{m}^2.
\label{cf}
\end{equation}
Here $C_0\approx 0.74$ is  the minimal value of $C$
and $f_m$ is the filling fraction for which 
$C(f)$ takes on its maximal value. $C(f)$ takes on its minimal value 
$C_0$ at the extreme filling fractions $f=0$ and $f=0.74$, 
and its maximal value is $C_m\approx 0.88$ at 
$f_m\approx 0.74/2$. The factor $0.14$ in the interpolation 
formula (\ref{cf}) is the difference $C_m-C_0$.
Let $k_L$ be the length of the Bloch vector at the $L$ point
of the Brillouin zone.
In units where the length of the side of the conventional 
unit cell of the cubic lattice is $A=2$, one has $k_L/\pi = \sqrt{0.75}$.
The formula
\begin{equation}
\triangle^r_L=2 \pi n_{e\!f\!f} \triangle_L/ k_L
\label{relg}
\end{equation}
then describes $\triangle^r_L$  for 
$1\leq \delta\leq 100$ with the relative error
ranging from $\approx 4\%$ for $f$  around $0.2$, to the relative error
$\approx 8\%$ for the close-packed case.
Formula (\ref{relg}) describes $\Delta_L^r$ 
in the air sphere case irrespective whether
there is ($f \geq 0.5$) or is not ($f < 0.5$)
a Mie resonance within the L-gap frequencies.
However, the formula is violated in the narrow interval 
$\delta\in(7.986,7.998)$ where the local widening of $\Delta_L^r$ 
takes place.

\section{Conclusion}
\label{sec:concl}
Using the photonic analog of the KKR method \cite{MS,Mo}, we have 
investigated Mie-resonance-induced effects for an ordered
medium, namely, for a simple fcc lattice  of  homogeneous spheres in 
three dimensions.  Our work partially fills the gap in the understanding 
of resonance-induced effects compared to a disordered medium \cite{vATL}.
We showed that the same principles apply 
to the classification of both electronic and electromagnetic bands
in a periodic medium, although their respective qualitative behaviour  
may be different. For example, due to the absence of bound
states for a single scatterer, it is much more difficult to open a
full gap in the spectrum of electromagnetic waves than for the
case of electrons. Still, under certain conditions, one can
identify a resonance band, and a hybridization of 
the Bragg and resonance bands can take place,
leaving behind a gap over the approximate
energy range of the original unhybridized resonance band \cite{Har}.
 We investigated both a partial hybridization
in the case of the first stop gap in the (111) crystal direction
(the L-gap) and a full hybridization in the case of the full band gap.
For dense spheres ($\varepsilon_s >\varepsilon_b$),
partial hybridization around the lowest Mie
resonance  occurs for all filling fractions $f$,
once the  dielectric contrast $\delta$ reaches a critical 
value $\delta_c(f)$ (see Figure \ref{rvff}). 
For air spheres ($\varepsilon_s < \varepsilon_b$),
the hybridization follows an irregular pattern and it can be observed 
only if $f\geq 0.5$. However, the value $\delta_c(f)$ can be 
as little as half the corresponding $\delta_c(f)$ in the dense-sphere case.
Near close packing ($f\approx 0.74$), 
the partial hybridization occurs around the second Mie resonance. 
The full hybridization in the case of air spheres
(there is no full gap in the dense sphere case \cite{SHI,BSS,MS})
follows again the irregular pattern seen in the case of 
partial hybridization. Sometimes the full gap opens
in a frequency region which does not correspond
to any Mie resonance ($f=0.68$),
 but in other situations there are up to three Mie resonances 
in the frequency range corresponding to the full gap 
($f=0.74$). The resonant-induced widening of a gap can occur if a Mie
resonance is about to cross the edge of the gap
(see Figure \ref{rlrgap}).
The values of the dielectric contrast required  to observe some of
these effects are within experimental reach
at microwaves and,  in the air-sphere case, 
even  at optical and near-infrared frequencies.
However, unlike in the case of the one dimensional Kronig-Penney model 
\cite{Jo}, we did not find any evidence that if some of Mie resonance
frequencies fall inside a gap, then this leads to its significant widening.
Contrary to the suggestions made previously in the literature
\cite{Jo}, no spectacular effects may be expected.
This is probably related to the known fact that the 
hybridization of bands in higher dimensions is weaker \cite{Har}. 
Also, since the opening of a gap in the spectrum of electromagnetic waves
in much difficult than in the electronic case, it is natural
that also hybridization effects are weaker in the 
photonic case.

\section{Acknowledgement}
We would like to thank A. van Blaaderen and
B. Noordam for careful reading of the 
manuscript and useful comments,
and members of the photonic crystals interest group for discussion.
This work is part of the research program of the Stichting voor 
Fundamenteel Onderzoek der Materie  (Foundation for Fundamental Research on
Matter) which was made possible by financial support from the Nederlandse
Organisatie voor Wetenschappelijk Onderzoek (Netherlands Organization for
Scientific Research).  SARA computer facilities are also gratefully
acknowledged.


\noindent
\newpage

\noindent
\end{document}